\documentclass{article}
\usepackage{spconf,amsmath,graphicx,bm,xspace,amssymb,amsfonts,textcomp,multirow,float,hyperref,algorithm,algpseudocode}
\usepackage{xcolor}
\long\def\comment#1{}

\newcommand{\fref}[1] {Fig.~\ref{#1}\xspace} % the non-breaking space "~" is very important!  use macros!

\newcommand{\xmath}[1] {\ensuremath{#1}\xspace}
\newcommand{\blmath}[1] {\xmath{\bm{#1}}}

\newcommand{\Var}{\mathrm{Var}}

\newcommand{\resp}[1]{}
\long\def\red#1{\bgroup\color{red}#1\egroup}

\newcommand{\etat} {\xmath{\eta_{t}}}

\newcommand{\argmin} {\operatornamewithlimits{arg\,min}} % argmin
\newcommand{\argmax} {\operatornamewithlimits{arg\,max}} % argmin

\newcommand{\bset}[1]{\xmath{\left\{#1\right\}}}

\newcommand{\complex} {\xmath{\mathbb{C}}}

\newcommand{\CN} {\xmath{\complex^N}}
\newcommand{\CM} {\xmath{\complex^M}}
\newcommand{\CMN} {\xmath{\complex^{M \times N}}}

\newcommand{\A} {\blmath{A}}

\newcommand{\T} {\blmath{T}}

\newcommand{\x} {\blmath{x}}
\newcommand{\xh} {\xmath{\hat{\x}}}
\newcommand{\xt} {\xmath{\tilde{\x}}}
\newcommand{\xfinal} {\xmath{\xt_{\scalebox{.8}{$\scriptscriptstyle \Nadd$}}}}

\newcommand{\y} {\blmath{y}}
\newcommand{\yh} {\xmath{\hat{\y}}}
\newcommand{\yk} {\xmath{\blmath{y}^{(k)}}}

\newcommand{\vveps} {\blmath{\varepsilon}}

\newcommand{\thta} {\blmath{\theta}}

\newcommand{\Nadd} {\xmath{N_{\mathrm{add}}}}
\newcommand{\Nstep} {\xmath{N_{\mathrm{step}}}}

\newcommand{\Nsmp} {\xmath{N_{\mathrm{sample}}}}

\long\def\red#1{\bgroup\color{red} #1 \egroup}
\def\x{{\mathbf x}}

\title{Adaptive Sampling for Linear Sensing Systems via Langevin Dynamics}
\name{Guanhua Wang, Douglas C. Noll, Jeffrey A. Fessler \thanks{This work is supported in part by
NIH Grants R01 EB023618 and U01 EB026977,
and NSF Grant IIS 1838179.}}
\address{University of Michigan\\Dept. Biomedical Engineering \& EECS\\
Ann Arbor, MI, 48105}
\begin{document}
%\ninept
%
\maketitle
\begin{abstract}
Adaptive or dynamic signal sampling in sensing systems can 
adapt subsequent sampling strategies 
based on acquired signals, 
thereby potentially improving image quality and speed.
This paper proposes a 
Bayesian method for adaptive sampling
based on greedy variance reduction
and stochastic gradient Langevin dynamics (SGLD).
The image priors involved can be either analytical or
neural network-based.
Notably,
the learned image priors generalize well 
to out-of-distribution test cases
that have different statistics than the training dataset.
As a real-world validation,
the method is applied to accelerate
the acquisition of magnetic resonance imaging (MRI).
Compared to non-adaptive sampling,
the proposed method effectively
improved the image quality 
by 2-3 dB in PSNR,
and improved the restoration 
of subtle details.
\end{abstract}
\section{Introduction}
\label{sec:intro}

Many imaging systems
acquire measurements sequentially.
Reducing the number of measurements
can accelerate the signal acquisition process
and benefit modalities that require lower radiation,
such as computed tomography (CT) 
and scanning electron microscopy (SEM).
Nevertheless, this can result in an under-determined image reconstruction problem.
To address this challenge,
various reconstruction methods have been proposed,
such as compressed sensing \cite{donoho:2006:CompressedSensing}, 
to enable the recovery of an object from undersampled measurements.

Sampling strategy also plays a critical role
in achieving high-quality images. 
For instance, many sub-Nyquist sampling patterns have been investigated in MRI,
including analytical and data-driven designs \cite{zibetti:2021:FastDatadrivenLearning}.
However, predetermined strategies may not always be optimal
for various imaging scenarios.
To address this challenge,
adaptive sampling or dynamic sampling techniques 
can select the next batch of `important' data points 
based on existing observations.
This approach enables better use of prior information
from both signal statistics and observed signals, 
leading to improved image quality and acquisition speed.
Relevant methods include Bayesian experimental design (BED) \cite{haldar:2019:OEDIPUSExperimentDesign},
neural network-based regression \cite{zhang:2019:ReducingUncertaintyUndersampled},
and reinforcement learning \cite{pineda:2020:ActiveMRKspace}.
These methods improved image quality in various applications.
However, many neural network-based methods may lack generalization ability and explainability to out-of-distribution
test sets and real-world applications.

This paper presents a model-based dynamic sampling approach 
that predicts new sampling locations
by greedily minimizing the variance of 
posterior samples drawn from the posterior distribution \cite{dilshangodaliyadda:2014:ModelbasedFrameworkFast}.
The sampler uses stochastic gradient Langevin dynamics (SGLD) \cite{welling:2011:BayesianLearningStochastic}
and supports various image priors.
We applied the proposed dynamic sampling
to accelerate MRI acquisition.
Across many experiment settings,
the proposed approach significantly improved the image quality.

\section{Methods}
\label{sec:methods}

\begin{figure*}[htbp]
    \centering
    \includegraphics[width=0.95\textwidth]{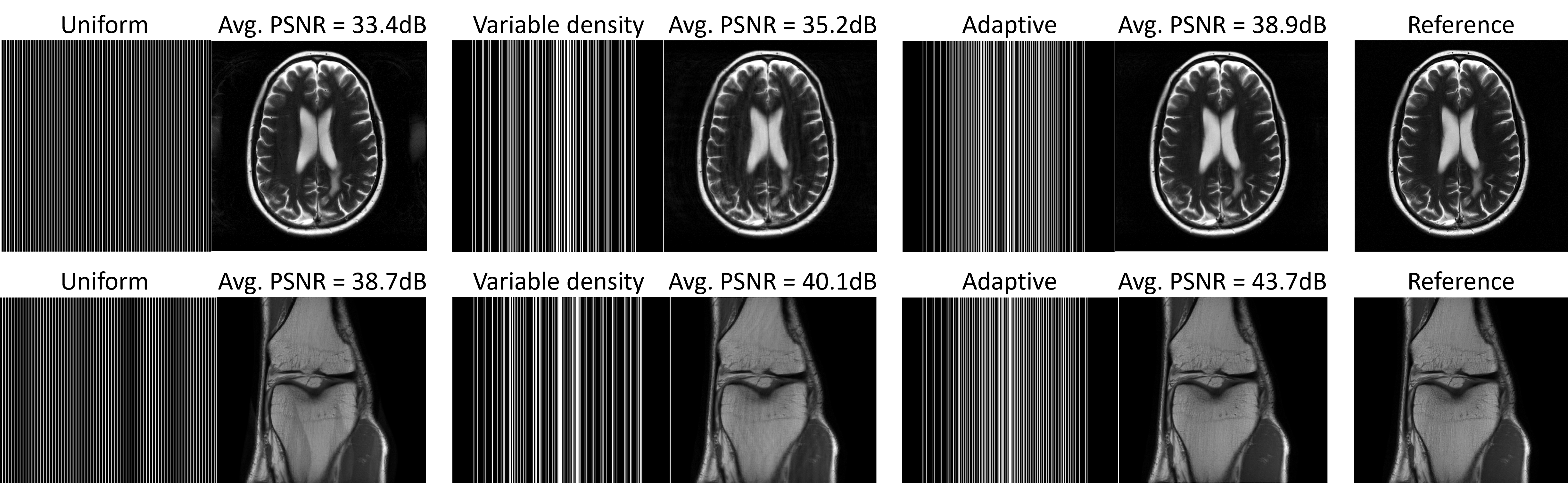}
    \caption{Comparison of different 1D sampling strategies with the analytical (roughness) prior. The undersampling ratio is 10$\times$ for all sampling patterns. The test set has $n$=20 slices. Dynamic sampling leads to reduced aliasing artifacts.}
    \label{fig,1d}
\end{figure*}
Consider a linear sensing system
\[
\y = \A \x + \vveps,
\]
where $\A \in \CMN$ denotes a sensing matrix,
$\x \in \CN$ denotes the object, 
and $\y \in \CM$ denotes raw measurements.
To accelerate the acquisition, 
we consider the `undersampled' case where
\y has $L<N$ non-zero entries.
Typically, the locations of non-zero entries in \y
follows pre-determined patterns.
The proposed method, instead,
dynamically chooses additional sample locations
in a sequence of $K$ sampling iterations
where the samples for iteration $k+1$
are based on the measurements
$(\y_1,\ldots,\y_k)$
recorded in previous iterations.

Specifically,
we apply a Bayesian approach
\cite{welling:2011:BayesianLearningStochastic}.
At the $k$ iteration of additive sampling,
based on
the measurements acquired up until this point
$\yk = (\y_1,\ldots,\y_k)$,
the first step draws samples from
the posterior distribution
$p(\x|\yk)$,
yielding a collection of reconstructed images
denoted $\{\xh_i^{(k)}\}_{i=1}^{\Nsmp}$.
We use an SGLD sampler detailed below.
The second step projects each estimate
$\xh_i^{(k)}$
(typically in the image domain)
back to the \textit{measurement domain}
using the sensing equation
$\yh_i^{(k)} = \A \xh_i^{(k)}$.
The third step selects the next sampling locations
by greedily minimizing the variance
of samples \bset{\yh_i^{(k)}}
in the measurement domain.
In detail,
we select
the next measurement location(s) $l$
for the $k+1$ iteration
using the k-space locations having the maximum variance:
\[
l =
\argmax_{n \in {1,2, \ldots N}} 
          \Var\{ [\yh_1^{(k)}]_n, \ldots, [\yh_{\Nsmp}^{(k)}]_n \}
.\]

To compute
a collection of reconstructions or estimates
$\{\xh_i\}$,
we sample from the posterior
\[
\xh \sim p(\x|\yk)
= p(\x) p(\yk|\x) / p(\yk) 
,\]
where $p(\x)$ denotes the prior
and $p(\yk|\x)$ denotes the likelihood.
In contrast,
a typical iterative image
reconstruction algorithm
gives a point estimate,
such as the MAP estimator.
SGLD \cite{welling:2011:BayesianLearningStochastic}
samples from the posterior distribution using the update
\[
\Delta \x_t = \etat (\nabla \log p(\yk|\x_t) + \nabla \log p(\x_t))
 + \sqrt{2 \etat} \, \mathcal{N}(0,\,1), 
\]
where \etat denotes 
the time-dependent step size \cite{hyvarinen:2005:EstimationNonNormalizedStatistical, song:2020:Generative}.
Intuitively,
SGLD explores the solution space by
injecting Gaussian noise
similar to the Langevin Monte Carlo sampler.

\algnewcommand\INPUT{\item[\textbf{Input:}]}%
\algnewcommand\Output{\item[\textbf{Output:}]}%
\algnewcommand\DATA{\item[\textbf{Data:}]}%

\begin{algorithm}[htbp]
  \caption{Adaptive sampling algorithm}
  \label{alg}
  \begin{algorithmic}[1]
    \Require Score function $f_{\thta}(\x) \approx \nabla \log p(\x)$;
    number of additive dynamic sampling iterations \Nadd;
    number of SGLD steps \Nstep;
    number of samples drawn from a posterior distribution \Nsmp;
    step size in SGLD $\blmath{\eta}$;
    penalty parameter for image prior $\blmath{\mu}$;

    \State Acquire initial measurements $\y^0$
    \State (optional) Pre-train $f_{\thta}(\x)$ on dataset $\mathcal{X}$
    via score matching.
    \For{$k$ = 1 to \Nadd} 
        \For{$i$ = 1 to \Nsmp}
            \For{$t$ = 1 to \Nstep}
            \State Initialize $\xt_0$
            \State $\xt_t = \xt_{t-1}+ \blmath{\mu}_{t} f_{\thta}(\xt_{t-1}) -
            \blmath{\mu}_{t} \blmath{\eta}_{t} \A'(\A\xt_{t-1} - \yk) 
            +\sqrt{2\blmath{\mu}_{t}} \, \mathcal{N}(0,\,1)$
            \EndFor
            \State $\xh_i^{(k)} = \xfinal$
            \State $\yh_i^{(k)} = \A \xh_i^{(k)} + \vveps$
        \EndFor
        \State $l = \argmax_{n \in {1,2, \ldots N}} 
          \Var\{ [\yh_1^{(k)}]_n, \ldots, [\yh_{\Nsmp}^{(k)}]_n \}$.
        \State Acquire additive measurements with index $l$
        and concatenate it
        with previous measurements $\yk = [\y^{(k-1)},~ y_l]$.
    \EndFor
\end{algorithmic}
\end{algorithm}
% \vspace{-1.5\baselineskip}

In applications where the noise \vveps is Gaussian,
the gradient of likelihood has the closed-form solution
$\nabla \log p(\y|\x) = -\A'(\A \x - \y).$
The prior term $p(\x)$,
or the score function $\nabla \log p(\x)$
can take various forms.
For example, a simple prior that
penalizes first-order roughness 
has the form
$p(\x) = \mathrm{e}^{-\lambda\|\T \x\|_2^2/2}$,
where \T is the first-order finite difference transform;
its corresponding score function is
$\nabla \log p(\x) = -\lambda\T'\T\x$.
Analytical priors may not
be informative
and many studies propose to
learn score functions from datasets.
Score matching 
approximates the score function
with a learnable function $f_{\thta}(\x)$ 
and learns from a training set $\mathcal{X}$:
\[
\argmin_{\thta} \mathbb{E}_{\x \in \mathcal{X}} \| \log p(\x) - f_{\thta}(\x) \|_2^2
\]

Recent improvements in score matching,
such as sliced score matching and
denoising score matching \cite{hyvarinen:2005:EstimationNonNormalizedStatistical, song:2020:Generative},
have extended the method's effectiveness 
and made it more applicable to large datasets \cite{ho:2020:Denoising,song2021scorebased}
To demonstrate the adaptability of
our algorithm, we tested
both analytical priors 
and score functions based on neural networks.
Alg. \ref{alg} details the proposed approach.

\begin{figure}[htbp]
    \centering
    \includegraphics[width=0.95\columnwidth]{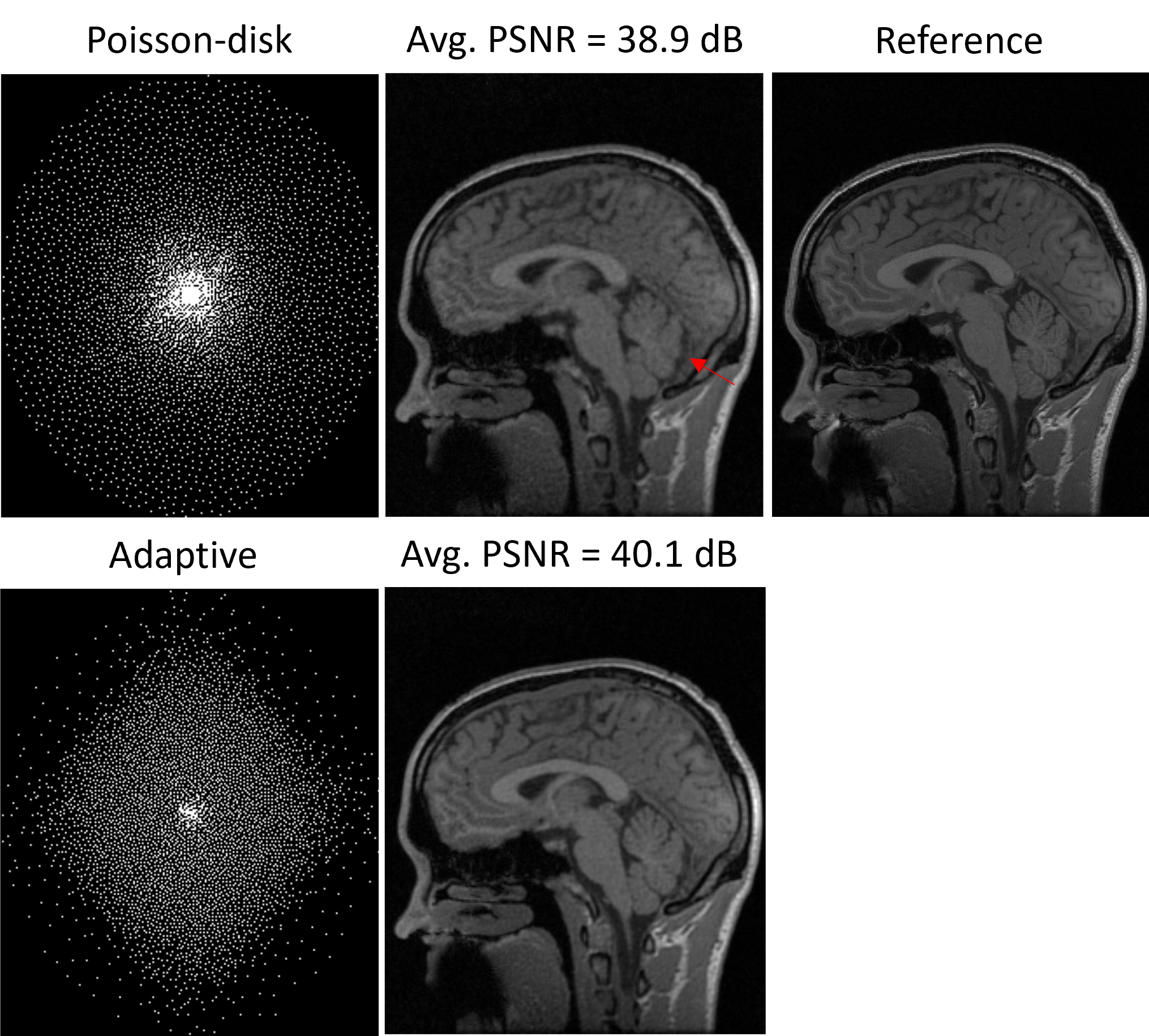}
    \caption{Comparison of 2D sampling strategies
    with reconstruction based on an analytical (roughness) prior.
    The undersampling ratio was 12$\times$ for both sampling patterns.
    The test set had $n=10$ volumes.
    Dynamic sampling reduced blurring and artifacts.}
    \label{fig,2d}
\end{figure}

\begin{figure}[htbp]
    \centering
    \includegraphics[width=0.95\columnwidth]{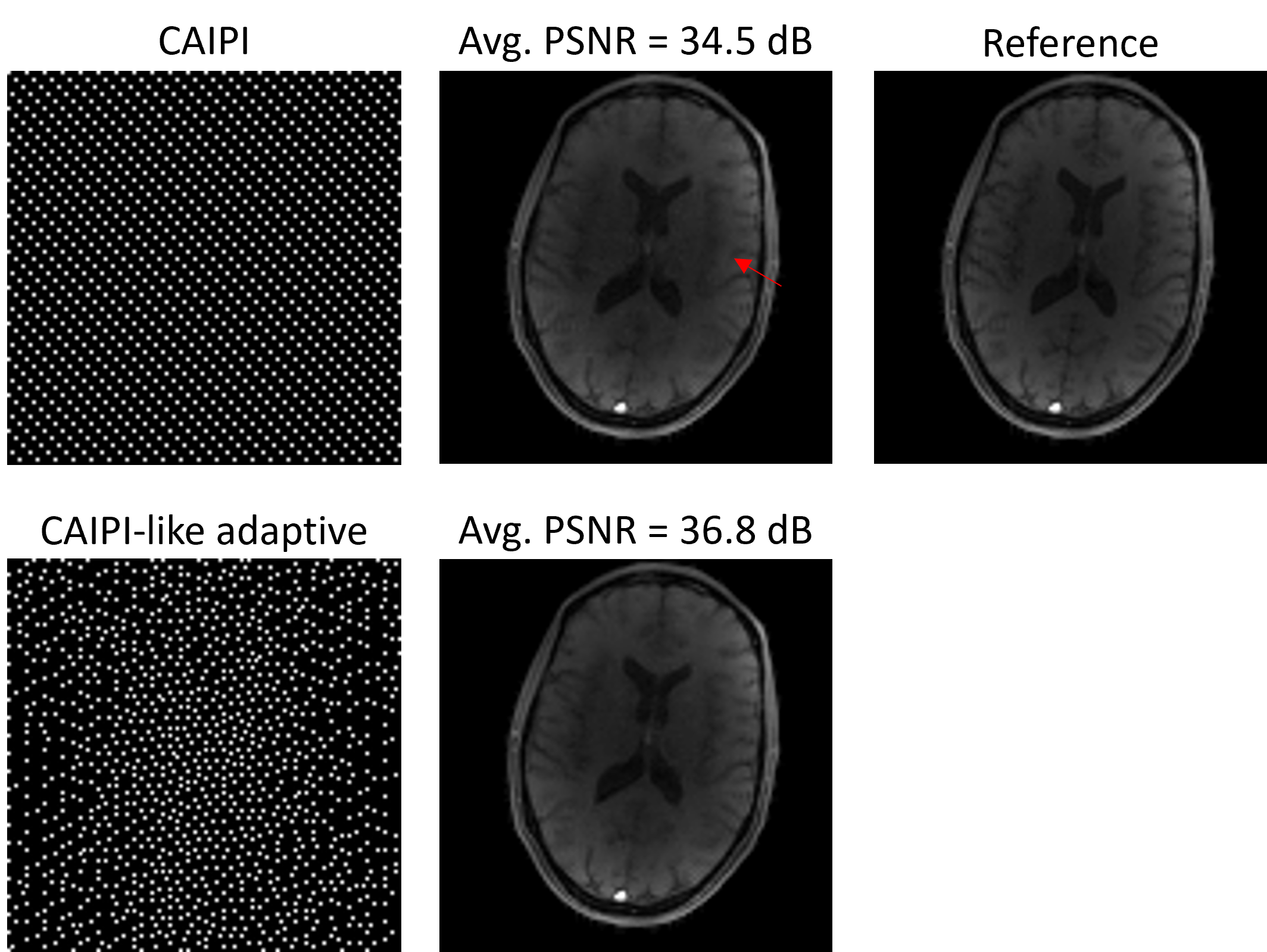}
    \caption{Comparison of 2D sampling strategies with the learned (NCSN++) prior.
    The undersampling ratio was $10\times$ for both sampling patterns.
    The test set had $n=16$ slices.
    Adaptive sampling improved tissue contrast and reduced blurring.}
    \label{fig,gre}
\end{figure}

\section{Experiments}
\label{sec:exp}

We applied the proposed dynamic sampling method to MRI data
that reside in the Fourier domain (k-space).
For our experiment with Cartesian sampling,
the sensing matrix \A contained both
FFT and coil sensitivity
(calculated by methods described \cite{ueckerESPIRiTEigenvalueApproach2014}).
The score functions included both
a simple analytical one $f(\x) = -\lambda\T'\T\x$
and a learned U-Net-based model.
We evaluated the analytical priors on multiple MRI datasets \cite{fastmri, desai2022skm,souza2018open},
using both 1D and 2D sampling patterns.
We compared the dynamic sampling patterns
with well-received fixed sampling patterns, such as Poisson-disk,
for
\Nadd = 50 and \Nstep = 200.

We used the same U-Net-based architecture (NCSN++)
and configurations as in \cite{song2021scorebased}
to train the learned prior
on the fastMRI brain dataset.
The complex-valued image
was formulated as two input channels.
To demonstrate the generalization ability,
we tested it on test sets
that contained different anatomies and sequences than the fastMRI database,
including an MP-RAGE sequence of human brains \cite{souza2018open} and 
a GRE sequence of mouse brains,
without any fine-tuning. 
For the mouse brain dynamic contrast-enhanced (DCE) data,
we learned the sampling pattern from a `pilot' frame 
and then applied it to subsequent frames.
We used
\Nadd = 30
and
\Nstep = 100
and the accelerated sampler described in \cite{chung:2022:ComeCloserDiffuseFasterAcceleratingConditional}.
The sequence $\blmath{\eta}$
used the same configuration as described in \cite{song2021scorebased}.

\section{Results}
\label{sec:res}
For the analytical prior,
\fref{fig,1d} and \fref{fig,2d} show
dynamic sampling patterns
and corresponding reconstruction examples.
Compared to predetermined sampling patterns, 
the proposed method reduced aliasing artifacts across multiple anatomies and contrasts.

For the learned prior (NSCN++),
\fref{fig,gre} shows an
out-of-distribution example, 
using GRE sequences of the human brain.
With the proposed adaptive sampling, 
the fine details and tissue contrast
in the reconstructed images
were improved
compared to predetermined sampling patterns.
\fref{fig,dce} shows another out-of-distribution case, mouse brain DCE imaging. 
The adaptive sampling scheme was optimized
for the first frame
and applied to subsequent frames.
Adaptive sampling led to less blurred structures
and improved SNR.

\section{Discussion}
\label{sec:diss}
The posterior sampling processes can be computationally expensive,
determined by both the system matrix \A
and the score function $\nabla \log p(\x)$.
Simpler analytical priors may accelerate the sampling.
The sampling is embarrassingly parallel
and can benefit from
parallel computing and hardware improvements.
In its current form,
the proposed dynamic sampling is particularly useful
for dynamic imaging applications such as fMRI and DCE-MRI
where a `pilot' scan is available to design tailored
sampling patterns for subsequent frames
and avoid the long computation
time that may compromise the benefits of dynamic sampling.

The sampling from the posterior distribution may benefit from faster
samplers \cite{xu:2022:Poisson}.
Some `single-shot' samplers based on neural network methods
can sample faster than SGLD \cite{tezcan:2022:SamplingPossibleReconstructions}
however, they are trained on a certain dataset and may
lack the ability to generalize to out-of-distribution applications.

The proposed dynamic sampling method has demonstrated decent robustness
in simulated experiments
and analytical priors worked well for different test cases.
The learned priors were trained on a fastMRI brain dataset
but generalized well to different anatomies, vendors,
sequences, and field strengths.
Future work will include a systematic comparison 
with prior arts and prospective in-vivo experiments.

% For many imaging applications,
% streaming the data in real-time is also challenging.
% For example, MRI may use frameworks from vendos such as Siemens FIRE.
\begin{figure}[htbp]
    \centering
    \includegraphics[width=0.95\columnwidth]{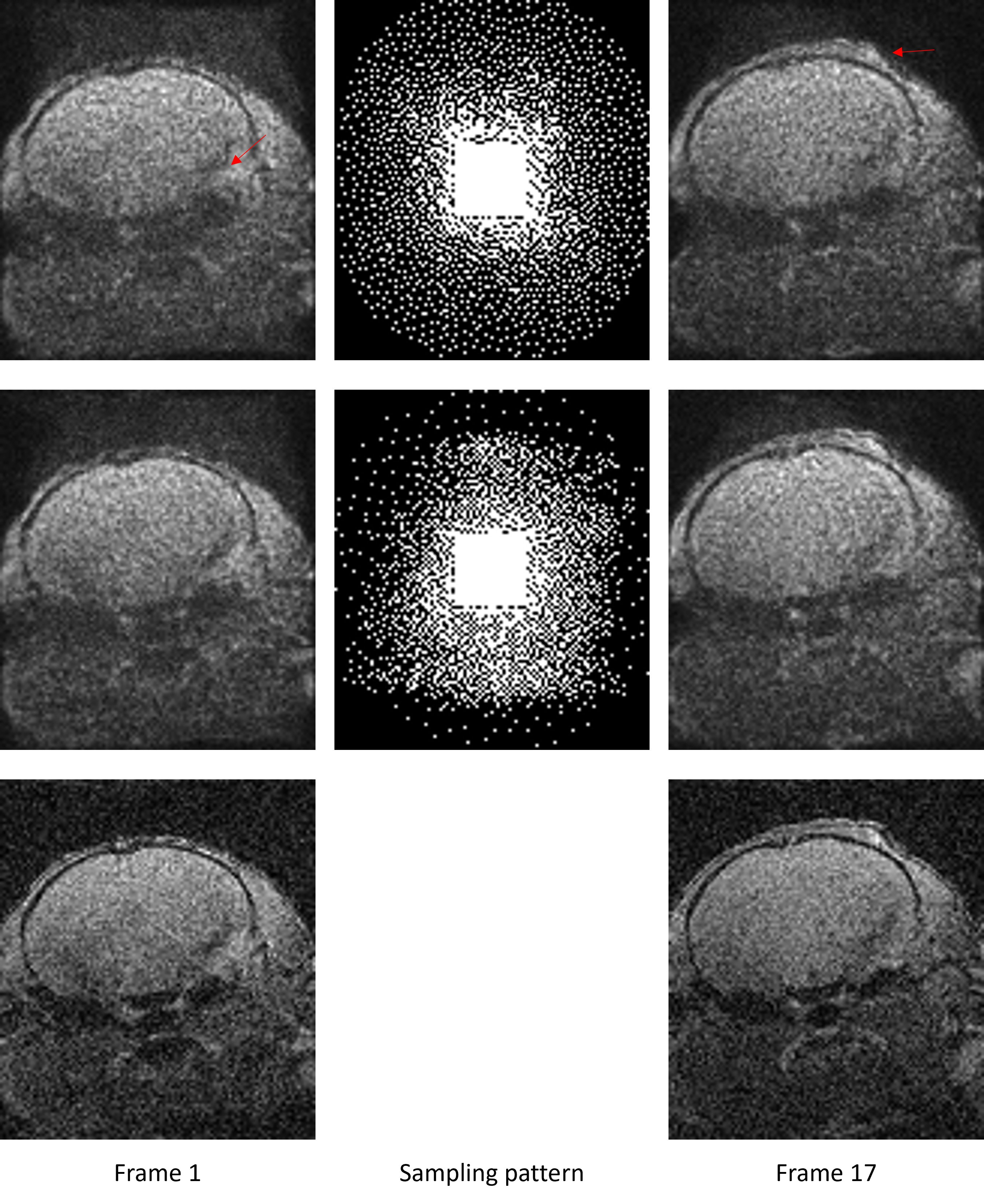}
    \caption{Comparison of 2D sampling strategies with the learned (NCSN++) prior. 
    The first row shows the Poisson-disk sampling pattern.
    The second row displays the adaptive sampling pattern optimized with the 1st frame and applied to the 17th frame.
    The third row shows the reference images.
    The undersampling ratio was 4$\times$ for both sampling patterns.
    Adaptive sampling led to reduced artifacts and higher SNR across different time frames.}
    \label{fig,dce}
\end{figure}

% References should be produced using the bibtex program from suitable
% BiBTeX files (here: strings, refs, manuals). The IEEEbib.bst bibliography
% style file from IEEE produces unsorted bibliography list.
% -------------------------------------------------------------------------
\bibliographystyle{IEEEtran}
\bibliography{refs}

\end{document}